# Replicability of Simulation Studies for the Investigation of Statistical Methods: The RepliSims Project

K. Luijken[1,2*], A. Lohmann[1*], U. Alter[3^], J. Claramunt Gonzalez[4^], F.J. Clouth[5,6^], J.L. Fossum[7,8^], L. Hesen[1^], A.H.J. Huizing[9^], J. Ketelaar[1^], A.K. Montoya[7^], L. Nab[1^], R.C.C. Nijman[1^], B.B.L. Penning de Vries[1,10^], T.D. Tibbe[7^], Y.A. Wang[11^], R.H.H. Groenwold[1,10^]

[1] Department of Clinical Epidemiology, Leiden University Medical Centre, Leiden, The Netherlands

[2] Department of Epidemiology, Julius Center for Health Sciences and Primary Care, University Medical Center Utrecht, University Utrecht, The Netherlands

[3] Department of Psychology, York University, Toronto, ON, Canada

[4] Methodology and Statistics Unit, Institute of Psychology, Leiden University, Leiden, The Netherlands

[5] Department of Methodology and Statistics, Tilburg University, Tilburg, The Netherlands

[6] Netherlands Comprehensive Cancer Organisation (IKNL), Utrecht, The Netherlands

[7] Department of Psychology, University of California, Los Angeles, Los Angeles, CA, United States of America

[8] Department of Psychology, Seattle Pacific University, Seattle, WA, USA

[9] TNO (Netherlands Organization for Applied Scientific Research), Expertise Group Child Health, Leiden, The Netherlands

[10] Department of Biomedical Data Sciences, Leiden University Medical Centre, Leiden, The Netherlands

[11] Department of Psychology, University of Toronto, Toronto, ON, Canada

* Both authors contributed equally.

^ Authors contributed equally.

**Abstract**

Results of simulation studies evaluating the performance of statistical methods are often considered actionable and thus can have a major impact on the way empirical research is implemented. However, so far there is limited evidence about the reproducibility and replicability of statistical simulation studies. Therefore, eight highly cited statistical simulation studies were selected, and their replicability was assessed by teams of replicators with formal training in quantitative methodology. The teams found relevant information in the original publications and used it to write simulation code with the aim of replicating the results. The primary outcome was the feasibility of replicability based on reported information in the original publications. Replicability varied greatly: Some original studies provided detailed information leading to almost perfect replication of results, whereas other studies did not provide enough information to implement any of the reported simulations. Replicators had to make choices regarding missing or ambiguous information in the original studies, error handling, and software environment. Factors facilitating replication included public availability of code, and descriptions of the data-generating procedure and methods in graphs, formulas, structured text, and publicly accessible additional resources such as technical reports. Replicability of statistical simulation studies was mainly impeded by lack of information and sustainability of information sources. Reproducibility could be achieved for simulation studies by providing open code and data as a supplement to the publication. Additionally, simulation studies should be transparently reported with all relevant information either in the research paper itself or in easily accessible supplementary material to allow for replicability.

*Keywords*

Replication, simulation studies, statistical methods, open materials

# Introduction

Many fields of empirical research rely on statistical estimation of quantitative phenomena. The value of the results of such studies depends, amongst other things, on the validity of the methods being used (Altman, 1994). Under strict assumptions and for relatively simple methods it is possible to mathematically derive how statistical methods will behave when applied to real data, e.g., whether type I error rates are correct or to what extent a method is able to identify an association if it truly exists (Boulesteix et al., 2017). However, for more complex research scenarios or more complex methods, the performance of a statistical method is usually assessed by means of statistical simulation studies (Burton et al., 2006; Morris et al., 2019). Simulation studies are computer experiments in which synthetic datasets are generated using computer algorithms (Boulesteix, Groenwold, et al., 2020; Burton et al., 2006; Morris et al., 2019). A key feature of these experiments is that the mechanism by which the data are generated is known and can, therefore, serve as a benchmark against which methods are compared. In addition, the flexibility of simulation studies in changing the data-generating mechanism means that methods can be tested under various conditions, such as different sample sizes, numbers of variables, and relations between variables.

Results of simulation studies often have a major impact on the way empirical research is done and analyzed. A striking example is the simulation study performed by Peduzzi and colleagues (cited > 7,900 times on Google Scholar, April 2023) on the sample size required to fit a logistic regression model, which is one of the most commonly used statistical models in the biomedical sciences (Peduzzi et al., 1996). This simulation study has had a major impact and even led to a widely used rule of thumb: the "one-in-ten rule." However, this rule could not be replicated in a replication study by van Smeden and colleagues (2016), suggesting that the results of this study might not be as generalizable as its high citation count might indicate.

Although simulation studies are a powerful tool for methodological research, results from those studies, as the example of the 'one-in-ten rule' illustrates, are not definitive. Like empirical results, results from simulation studies need to be reproduced and replicated to verify their veracity (Boulesteix et al., 2017; Lohmann et al., 2021), and this is increasingly called for (Boulesteix, Hoffmann, et al., 2020). So far there is limited evidence on the reproducibility or replicability of simulation studies. Therefore, we aimed to investigate the extent to which highly cited simulation studies could be replicated. The present study neither sought to improve upon nor criticize the original authors' approaches.

**Reproducibility and replicability of simulation studies**

There is no broad agreement on what the term replicability means in the context of simulation studies (Claerbout & Karrenbach, 1992; Patil et al., 2016; Plesser, 2018; Rougier et al., 2017). For the purposes of this work, we rely on terminology defined in *The Turing Way* (The Turing Way, 2020) and extend it to consider the defining characteristics of reproduction and replication in simulation studies.

Reproducibility is defined as generating the exact same results using the exact same data and the exact same analysis (The Turing Way, 2020). Reproducibility in empirical research might look like applying analysis scripts which are publically posted to analyze (publically) available data to evaluate if results are the same as what is presented in the published paper. All research should, as a bare minimum, be reproducible, and failures to reproduce the results of a study suggest there could be an error or some other issue with the study that would reduce its value (Nosek et al., 2022). However, successful reproduction of a study does not add additional evidential weight (Goodman et al., 2016). Some have suggested that studies which are not reproducible should not be considered as candidates for replication, because the results of such a replication would be difficult to interpret in the absence of reproducibility (Nuijten et al., 2018). Reproduction attempts with empirical research have varied in their success rate, but none have been completely successful (Artner et al., 2021; Bakker & Wicherts, 2011; Hardwicke et al., 2021; Hardwicke et al., 2018; Maassen et al., 2020; Nuijten et al., 2016; Obels et al., 2020; Wang et al., 2022).

In the context of simulation studies, we believe it is important to extend the definition of reproducibility slightly from considering the exact same data to include the exact same *data generating process*[1](Bollen et al., 2015). For example, if open code for a simulation study is available, but the original study did not set the random seed as part of the analysis, then the exact same data cannot be recovered, but we believe that such a case falls squarely within the purpose of reproducibility. To consider an equivalent case for empirical research, there are many analysis strategies which rely on random number generation (e.g., bootstrapping, EM algorithms, multiple imputation), so considering a study with open data and open code but with no seed set for the analysis, the exact same results may not occur because of the randomness in the analysis. Still, though, we believe that this process fits the purpose of reproducibility. Reproducibility should be a minimum standard for simulation studies: Providing open data and code poses no ethical barriers and thus should be required for all published simulation studies.

Replicability for empirical data is defined as conducting the same analysis with different data collected using methods as similar as possible to the original study and obtaining a similar result (The Turing Way, 2020). Even in an ideal world, we would not expect all research to replicate, because type I and type II errors are probablistically defined. A failure to replicate the results of a study might call into question the broader theory supported by the findings (Nosek et al., 2022). Alternatively, successful replication would provide additional evidential weight to the claims from the original study (Goodman et al., 2015). Importantly for simulation studies, a replication would not involve the same data as the original study or (as described above) the same data generating process. Instead, the data generating process and the analysis would be independently created by the replicator based on the available information from the publication. A failure to replicate a simulation study might suggest that there were specific choices in the original study which influence the conclusions of the study but were not reported.

---

[1] This definition is in line with Bollen et al., (2015) which focuses on the same "materials" rather than the same data being used for reproduction.

## Methods

In this study, we focus on process replicability, meaning assessment of whether the original description of the simulation was understandable to the replicator(s) and the degree to which they were able to implement it. The focus of our work is on feasibility of recreating experimental conditions corresponding to the original studies. Equivalence of results found between the original study and the replication was used as a means to assess process replicability, i.e., we assume that results turn out similar if simulated data, implemented computations, and software functionalities are similar (Clemens, 2017; Goodman et al., 2016). We believe this is a reasonable assumption for computational research and thus that the focus on feasibility is relevant (assessment of feasibility is explained further below).

### Selection of studies

To investigate the replicability of highly cited statistical simulation studies, we identified eight studies that assessed the performance of statistical methods and are commonly cited within the field of health science or social science. These studies were identified by AL, AKM, and RHHG and are described in Table 1. We chose to focus on studies published after 2000 with a relatively high citation count, because these studies arguably have the largest impact on subsequent empirical studies. The number of citations for the included studies ranged from 1650 to 7098, based on Google Scholar citations retrieved in March 2022. The included studies were superficially scanned to ensure the included methods were of such a general nature that the replicators could be expected to complete the replication attempt, provided that sufficient information was reported. Notably, none of the original simulation studies provided open data or code, and so assessing reproducibility was not feasible.

**Replication set-up**

Teams of replicators retrieved relevant information for the replication from the original publication of their choosing. This information was then used to write simulation code in an open-source programming environment of choice with the aim of assessing the feasibility of recreating the experimental conditions which generated the original results. Results of the replication were compared to those reported in the original publication, with the primary outcome of our study being the feasibility of translating the information provided in the original studies into computer code.

***Replication teams***

Each study was replicated by teams of at least two replicators, consisting of a primary replicator and co-pilot(s). All replicators had formal training in quantitative methodology corresponding to the minimum of a M.Sc. degree in statistics, psychology, or epidemiology. All replicators had prior experience in conducting simulation studies. Replicators extracted information pertaining to the implementation of the simulation studies from the original publication and translated this information into simulation code. The primary replicator coded and ran the replication simulation. The co-pilot studied the simulation code and provided feedback as needed. If feasible, the simulation was run, and results were reported. If not feasible, we report barriers to replication.

### Information about the simulation studies

Relevant information for the replications was obtained from the original publications. Information that was explicitly referenced in a publication was also considered. Each team of replicators kept track of information that was ambiguously reported and noted assumptions that they had to make.

### Software

The replicators could choose any open-source programming environment for the replication irrespective of the original implementation. All replicators conducted their replications in the R statistical software (R Core Team, 2013). Details regarding corresponding packages and software versions can be obtained from the individual replication reports (provided in the Supplementary Materials). For reproducibility of our work, all replication code can be obtained from the project's GitHub organization https://github.com/replisims/.

### Assessment of replicability

Each replicator team aimed to replicate the original simulation study by creating simulation code and performing analyses as similar as possible to the original study based on the information provided in the manuscript. As the replication of simulation studies is a novel endeavor, there are currently no set criteria to assess the alignment of replicated simulation results with the original results. All factors hindering or facilitating the process were documented by the replicators, because these factors can provide valuable insights for the improvement of future simulation studies. Agreement between results from the replication studies and the original studies was assessed in a qualitative manner and involved evaluating whether numerical values from the replication studies were comparable to those in the original studies, whether trends in the results were moving in the same direction, and whether the performance rankings of different simulation scenarios matched those in the original studies. Replicators did not check for appropriateness of applied methods or correctness of the original methods.

While we focused on the information provided in the original publication, the original authors were contacted after the replication attempt was finished as a means to assess the accessibility of possible additional information that could facilitate a replication attempt. The original authors were not contacted earlier to eliminate the possibility of author-provided information influencing the interpretation of the original manuscript. The authors of each publication were contacted via email with a request for additional information or computer code pertaining to their simulation study. In case of non-response, a single reminder was sent. The original authors were not contacted until after the replication attempt was finished. This was done to eliminate any risk of author-provided information influencing the interpretation of the original manuscript.

## Results

We begin the results section with an overall summary of how feasible replication of each study was. Given the low number of replicated studies, we deemed a quantification of findings inappropriate. Instead, we identified features hindering or facilitating replicability of simulation studies by providing examples of the replicator's experiences. The discussion of experiences is narrative rather than systematic, meaning that examples listed are illustrative and not comprehensive. An overview of key aspects of replicability per study can be found in Table 2. In the Supplementary Materials, we provide an overview of individual study features which hinder and facilitate replication as well as replicator degrees of freedom, which we define as the flexibility involved in the process of replicating a (simulation) study.

**Overall feasibility of replicability**

In two studies, almost perfect replication of results was achieved (Brookhart et al., 2006; Fritz & MacKinnon, 2007). For one study, not enough information could be obtained to implement any of the reported simulation scenarios (Vittinghoff & McCulloch, 2007).

Replication was partially feasible in five studies. In the replication of the study by Austin (2011), parameter values reported in the paper did not align with the description of the properties of the data of the original study. Therefore, it was unclear whether the implemented mechanism was in line with the original simulation. The replication of Flora and Curran (2004) as well as Rhemtulla and colleagues (2012) led to results that were overall consistent with the original simulation results. For Flora and Curran (2004), there were differences between the replication study and the original study in the rates of improper solutions and the direction and magnitude of relative bias of the factor loadings, possibly due to the use of different software environments. That is, the described implementation of the statistical method in the original publication could not always be replicated exactly. In the replication of Rhemtulla and colleagues (2012), dissimilar results mostly occurred in scenarios in which the confirmatory factor analysis model did not converge or had a negative variance. Similar to the original study, scenarios with such errors were excluded from further analysis. The detailed descriptions of error frequency in the original study, made it possible to detect that scenarios with large discrepancies from the original study corresponded to scenarios with high numbers of errors. Given the large number of errors (also encountered in the original study), it would have been advisable to report Monte Carlo errors to allow a more nuanced comparison of the magnitude of discrepancies.

In the replication of the study by MacKinnon and colleagues (2004), the overall conclusions of the original article were replicated. The original article compared nine methods of constructing confidence intervals for the indirect effect in mediation analysis in terms of performance measures like power and type I error rates. In the replication study, the relative performance of these methods largely agreed with the original simulations, with a few exceptions: For four of the methods, and in particular the $M$ method and empirical-$M$ method, the sources provided in the original article were insufficient for the team to replicate the simulations. For instance, critical values used in one of the inferential methods had to be obtained from a book, but it was unclear how they were implemented in the original simulation study. As such, more decisions had to be made by the replicators about how to implement them, and one method (the empirical-$M$ method) had to be excluded from the simulation altogether because no way forward was found. The replication of Peters and colleagues (2006) yielded results that resembled the general pattern and direction of the original results. However, only part of the simulation results was presented in the original study's main text. Matching the replicated results to the displayed result was challenging, particularly because results were presented as figures only.

**Replicability hindering properties**

*Missing information*

When simulation parameters were insufficiently described, replication was hampered at an early stage. This was the case in the replication of the study by Vittinghof and McCulloch (2007). Some parameter values, such as the variance of predictor variables, were missing, and reporting on the selection of scenarios was incomplete. As a result, it was infeasible to replicate the set of parameters used to generate simulated data. In an attempt to recreate the reported set-up, the reported information led the replicators to specify 10,176 scenarios in the first simulation experiment, whereas the original study mentioned 9,328 scenarios only. Similarly, in the second simulation experiment, the reported information led the replicators to specify 4,240 scenarios rather than the 3,392 reported scenarios (see Supplementary File, Vittinghof and McCulloch replication report, p. 6-7 for ambiguities in the simulation parameters). The ability to verify the agreement of the replicated simulation with the original study was so low that replication was discontinued.

Sometimes other documents that were referred to for information could not be retrieved. For instance, in Flora and Curran (2004), links to the technical appendix and the data generation and analysis code from the published paper were broken[2], resulting in uncertainties about information not explicitly reported in the original paper (e.g., tau values used in data generation; see Table 2 for details). Broken links also made it difficult to implement several methods included in MacKinnon and colleagues (2004) and Fritz and MacKinnon (2007). Web addresses given in the original publication of MacKinnon and colleagues (2004) were supposed to connect to an algorithm and critical values needed to perform two of the methods, but one link no longer worked and the other led to a website that had since been updated and no longer contained the necessary critical values. As a result, the method requiring knowledge of these critical values had to be omitted from the replication simulation. Finally, the study by Peters and colleagues (2006) referred to a technical report that was not publicly available (the report was eventually retrieved after correspondence with the Department of Health Sciences at Leicester University, where the original study was conducted).

---

[2] After completion of the replication, it was discovered that the link from the publishers website (but not in the published paper) was functional, and the partial code could be recovered, but this code was not used in the replication attempt.

### Error handling

Lacking information on checking and handling of runs with nonconverged or inadmissible solutions (e.g., solutions with negative variance estimates) was another barrier to replication. In the replication of Flora and Curran (2004), the rate of nonconvergence was higher in some of the conditions than that reported in the original study. This was the case for confirmatory factor analysis models estimated using weighted least squares in settings of small sample sizes, i.e., 100 or 200 observations per simulated data set (see Supplementary File, Flora and Curran replication report, p. 18 for a full discussion of nonconvergence rates). Because fit statistics and parameter estimates could not be obtained from nonconverged models, these conditions were excluded from the replication. In replicating MacKinnon and colleagues (2004), a function not used in the original simulation was implemented to calculate one of the confidence interval methods (Tofighi & MacKinnon, 2011), because no alternative way of implementing the method could be gleaned from the original paper. This function produced errors under certain conditions, but the original paper did not discuss whether similar cases were encountered in the original simulation or what was done in such cases. Ultimately, the replicator team decided to rerun those cases which resulted in 13,264 rerun iterations (being 1.6% of the total number of iterations in the simulation experiments). Finally, the study by Rhemtulla and colleagues (2012) involved the simulation of continuous data from a confirmatory factor analysis which was subsequently categorized. Categorization could result in the sample covariance matrix being non-positive definite due to a perfect positive correlation between two or more variables or due to one or more variables having zero variance. While the manuscript did specify how several types of errors were handled, it was unclear how this particular issue was addressed.

### *Ambiguous information*

When studies referred to different sources for information, this information could not always be mapped back to the study. For instance, the study by MacKinnon and colleagues (2004) referred to a table in a book for the critical values used in one of the methods they examined in their simulation experiments, yet no further information was provided on how the values in the table were translated for use in the simulation procedure (e.g., how cases that resulted in values not exactly reported in the table were interpolated). Furthermore, the authors provided a citation for calculating the skewness of a distribution, but it was not clear from the simulation study and the cited text which distribution they were referring to.

Replication of the data-generating mechanism was impeded when data-generating procedures only contained a description of expected results without specifying the process to generate those results. For example, the study by Vittinghof and McCulloch (2007) contained a description of the expected prevalence of a binary predictor and its correlation with continuous predictors in the dataset, but it was not indicated how the binary predictor data was generated or how the correlation with other variables was introduced.

Replication was also impeded when important information was provided for some methods but not others. For example, Fritz and MacKinnon (2007) provided a margin of error on power level of 0.1% for methods that were evaluated in 100,000 simulation iterations; however, the margin of error was not explicitly stated for the methods that were evaluated in 1,000 simulation iterations. Applying the margin of error on power of 0.1% was infeasible where no sample size was found that had power within the narrow margin of error (e.g., a sample size of 34 could have 79.9% power and 35 could have 80.1% power, and neither are within the margin of error). The replicators ultimately chose to increase the margin of error to 0.5%. Notably, the largest numeric deviations between the original and replication studies were for the methods evaluated under the adjusted margin of error on power levels.

Sometimes information in different parts of the manuscript contradicted each other. For example, in the study by Peters and colleagues (2006), an effect-size based method to emulate publication bias was described. On the one hand it was suggested that extreme studies with a negative effect of the exposure should be censored. On the other hand, it suggested censoring 40% of studies, which seems unlikely with large effect sizes (e.g., an odds ratio of 5). Hence, these instructions could not be followed together. Similarly, Austin (2011) specified each coefficient for the data generating model. However, implementing the coefficients as specified did not result in the marginal probabilities implied in the original manuscript.

### Discrepancies in software

When a study used proprietary software, as was the case for Flora and Curran (2004), Fritz and MacKinnon (2007), and MacKinnon and colleagues (2004), replicability was hindered if this software could not be accessed by replicators. In replicating results by Flora and Curran (2004), the use of different (open source) software might have introduced differences in the number of improper solutions at small sample sizes and resulted in slightly different directions and magnitudes of some relative bias findings compared to the original study, because the software's default settings or computational strategies might have differed.

## Replicability facilitating properties

Although one could simply conceptualize replicability facilitating factors as abstaining from all the practices we described in the previous section, we would like to highlight specific features, which we found made our replication attempts easier and potentially more accurate.

### *Extensive documentation*

Extensive documentation made it easy to understand how the simulation experiment was set up in the original study and thereby facilitated replication. Journals often have limited space for such documentation, but a way to share extensive documentation is to present the information elsewhere, as was done in the technical report that accompanied the study by Peters and colleagues (2006). The study by Brookhart and colleagues (2006) provided formulas for the approaches studied as well as a depiction of the data-generating mechanism in a figure. These aspects provided clear guidance for how to set up the simulation experiment and made replication of the study relatively easy.

One example of well-structured documentation was provided by Rhemtulla and colleagues (2012). Information about each aspect of the simulation set up could be easily retrieved from the manuscript. Other examples where the overall structure of the simulation was easy to get from the original article, and where the simulation conditions were explicitly laid out, were Brookhart and colleagues (2006), Fritz and MacKinnon (2007), and MacKinnon and colleagues (2004).

*Availability of suitable software implementation*

Availability of (parts of) the simulation code clearly facilitates replication attempts. For example, for the study by Flora and Curran (2004), part of the simulation code was available as part of the SimDesign package in the R statistical software (Chalmers & Adkins, 2020), and this code was generalized for the replication. The methods investigated by Rhemtulla and colleagues (2012) were conducted using proprietary software in the original study; however, in the replication, it was possible to use the lavaan package (Rosseel, 2012), which is complemented by an entire structural equation modeling infrastructure for simulation studies (e.g. simsem, Pornprasertmanit, Miller & Schoemann, 2021). While this package did not provide any code of the original simulation, and the infrastructure facilitated the implementation of the methods.

*Clear presentation of findings*

Presentation of simulation results in tables rather than figures facilitated assessment of agreement of findings and was done by MacKinnon and colleagues (2004) and by Brookhart and colleagues (2006) for their first simulation experiment. We do not wish to suggest that for replicability purposes all figures in simulation studies should be replaced by tables. Rather, the approach by Rhemtulla and colleagues (2012) could be taken, where complete results were presented in table form in the supplemental materials.

**Discussion**

The present study attempted the replication of eight highly cited simulation studies investigating the performance of data analytical methods. No complete simulation code was openly available for the identified simulation studies, and so reproducibility was not investigated. In three studies, almost perfect replication of results was achieved (Brookhart et al., 2006; Fritz & MacKinnon, 2007; Rhemtulla et al., 2012). Replication was partially feasible in four studies (Austin, 2011; Flora & Curran, 2004; MacKinnon et al., 2004; Peters et al., 2006). For one study, not enough information could be obtained to implement any of the reported simulation scenarios (Vittinghoff & McCulloch, 2007). To the best of our knowledge, this is the first attempt to replicate a set of simulation studies and to provide a formal assessment of factors hindering and facilitating replicability.

Information provided in the original publication (plus accompanying documents) was not always sufficient for replication. This observation is in line with the results of a review of reporting practices of simulation studies (Morris et al., 2019). This observation is not unique to simulation studies and has been found in empirical research in the fields of medical and social sciences (Vachon et al., 2021). We speculate that this is partly due to certain details being considered trivial information by the original researchers (and reviewers). In case of space restrictions imposed at many journals, what is considered trivial or obvious may not be reported in detail. However, for successful replication by researchers not involved in the original research, a detailed description of the simulation procedure is essential, otherwise the replicator has to make (arbitrary) decisions which may be a source of discrepancy between results of the original simulation study and its replication. Those arbitrary decisions are part of the 'replicator degrees of freedom'. What is more, some of the replication studies were performed in a different programming language, which might have different default settings. This illustrates that decisions are sometimes made implicitly but might deserve explicit reporting.

The current work focused on replication of the simulation studies, meaning that we focused on whether similar results could be obtained if data generation and analysis were performed as similar as possible to the original study. For simulation studies, it would also be particularly relevant to assess the generalizability of findings about a method by exploring alternative approaches to testing the same question or evaluating novel conditions. Evaluation of when and how a method is ideally implemented requires a different type of methodological research than developing a new method (Heinze et al., 2022).

Several potential limitations of this study need to be addressed. The original simulation studies chosen for replication in the current study were selected based on topic, number of citations, and expertise of the replicators, and are not likely to be representative of simulation studies in general. Each replication team selected their own simulation studies to replicate as well, which could have led to their particular skill sets and interests influencing the studies they chose. With merely eight simulation studies being replicated, our sample was relatively small. Nevertheless, it provided valuable insights into factors that facilitate or hinder replicability of simulation studies. Also, although the replicators were formally trained in quantitative methodology and experienced in conducting simulation studies, they were not necessarily experts on the exact topics that were investigated in the original simulation studies. Possibly, tacit knowledge about a particular field or method could have enhanced replicability. For instance, the simulations by Fritz and MacKinnon (2007) and MacKinnon and colleagues (2004) were replicated by researchers who specialize in mediation analysis, and these were two of the more replicable studies.

Similarity of results was used to assess replicability. However, when a replication attempt yielded results similar to the original study, few or no further checks were conducted to see if implementation was actually similar, whether results were obtained due to coincidence, or whether errors were made but canceled each other out. In contrast, when results differed from those reported in the original study, the code was scrutinized, and some replication teams programmed several implementations to obtain the original result. In case of insufficient information being available, replicators had to make informed guesses about, for example, possible values of simulation parameters, since computationally intensive procedures prevent a trial-and-error approach to replication. Finally, it is worth noting that the current replication did not address the design of the simulation study itself, that is, how the original authors operationalized the research question.

The teams sought to replicate, rather than reproduce, results of the original study, meaning that we did not seek out original code and data from the research teams from the onset of the study. It may be worth considering whether reproduction should be a prerequisite for replication of simulation studies; however, none of the eight identified studies had openly available data or code (except Flora & Curran (2004) which had broken links), and so evaluation of reproducibility of these studies may not be possible. Making simulation code publicly available is not common practice yet and therefore we focused on what was reported in the original publications. To illustrate alternative routes to original code, we contacted the corresponding authors of the original studies by email after the attempts to replicate the simulation studies based on the information provided in the original publication were complete. All corresponding authors responded to our emails. In some situations, this led to additional information, including even the code used for four of the original simulations. At this point, reproducibility could have been evaluated for those studies, but because the focus of this study was on replicability, this information was not used to improve the replication and hence contacting the original authors did not have consequences for the replication attempts. Clearly, complementing the publication of a simulation study with (publicly) available simulation code would greatly enhance reproducibility and replicability. Journals which publish simulation studies should consider requiring code and data to be publicly available, similar to recent pushes in empirical research (Easterbrook, 2014; Stodden, 2010). Future research could consider identifying simulation studies with open code to evaluate reproducibility separate from replicability, as these two characteristics can speak to different properties of the original studies.

**Reproducibility and replicability of simulation studies**

Reproducibility and replicability of simulations studies are related but distinct goals, just as they are in empirical research. A replicable study is one where an independent research team could collect a new sample from a similar population using methods as close as possible to the original study, conduct analyses as similar as possible to the original study, and find a similar result. A failure to replicate could occur for many reasons, but a common concern is that there may be specific details of the original study that are not reported in the original manuscript or supplemental materials, yet are key to producing the same results. However, in empirical research, there is no evidence that input from original authors impacts the success rate of replication (Ebersole et al., 2020). A similar concern could arise for a simulation study, where key details about the data generation, data analysis, or aggregation are omitted from the available materials such that another reasonably expert research could not reproduce their results. A failure of replicability could have implications for whether the scientific community deams the research to be true and accurate.

Alternatively, reproducibility is more closely connected to the use of the original materials (code and/or data). Failure of reproduction, especially if attempts at reproduction result in differing conclusions, means that even given the original ingredients and recipe, identical results cannot be reproduced. Previous researchers have differentiated between process and outcome reproducibility (Nosek et al., 2022). A failure of process reproducibility refers to a lack of information available to generate the same results whereas a failure of outcome reproducibility means that all the materials are available, and still the same results do not occur. Both types of reproducibility could occur for simulation studies and we believe that reproducibility should be a minimal standard for simulation; however, while data and code sharing for simulations may seem obvious, there may be case-specific limitations which need to be considered (e.g., open code for complex simulations on super computers; what level of detail should be included in open data; complications around using alternative operating systems).

An obvious way of facilitating reproducibility of simulation studies would be to provide access to simulation code and data, which should then include details regarding simulation parameter settings, coding environments and dependencies (including their versions), random number generator seeds, and implementation of algorithms for data generation as well as data analysis and presentation of results. These details should be provided regardless of the software being used, although the use of open-source software would lower the barrier for reproduction.

While we perceive reproducibility to be a minimal standard, we believe that replicability should still be sought after in order to identify important details required for replication. A simulation study may be reproducible, but if an expert researcher cannot generate similar results based on the information reported in the paper and supplemental materials, this suggests that perhaps the results are subject to very specific decisions which were not reported in the manuscript. This issue is difficult because not every detail of a simulation study can be reported in a manuscript given length limitations; however, a standard to strive for is reporting any specific decisions that the research believes are tantamount to replicability. This type of description may be aided by including "Constraints on Generality" statements (Simons et al., 2017) in simulation studies.

To facilitate both reproducibility and replicability, transparent and clear information on how the study was conducted should be provided, and may be expressed in either code or manuscript text. Existing guidance by Morris and colleagues (2019) outlines how to report on main aspects, such as the aim of the simulation study, data-generating mechanism, estimand, methods, and performance measures. Additionally, reporting software-specific features such as definitions of improper solutions and version numbers facilitates assessment of reproduction. As indicated above, what is critical information and what is implicitly considered background knowledge may be hard to assess for researchers themselves. Therefore, making simulation code publicly available, e.g., in a repository or as an online supplement on a journal's website, is an important recommendation in order to improve replicability and reproducibility of simulation studies. An example of preparing code for peer-review is the Checklist for Code and Data Supplements from the Biometrical Journal (Hofner, 2015; Hofner et al., 2016).

Future studies on reproducibility and replicability of simulation studies are encouraged. The time needed to complete a replication differed per study and was not recorded but was estimated to be at least forty hours per replication. Similar to Nuijten et al. (2018), we believe that evaluation of reproducibility of simulation studies may be a useful first step prior to conducting a time and resource intensive replication; as a failed replication of a study that lacks reproducibility adds limited information to the scientific community. For future replicators, we recommend replicating a simulation study that is closely related to a planned research project, to undertake as a foundation for the study. This effort could be extended by investigating the robustness of findings under different data-generating mechanisms or implementation of approaches, i.e., to evaluate generalizability of findings. Finally, replication of simulation studies could be an educational project for trainees.

**Conclusions**

Discussions about replicability of research in the fields of biomedical and social sciences have focused on studies with human participants, where replicability may be impaired by heterogeneity of participants across studies. Such heterogeneity should not affect simulation studies investigating statistical methods, which therefore should be perfectly or near-perfectly replicable. This pioneering study showed, however, that replicability of simulation studies is not a given, and the information provided in the original publication of highly cited and influential simulation studies was often insufficient for complete replication. We encourage researchers who publish simulation studies to transparently report all relevant information and preferably make their simulation code and data publicly available to facilitate future research, including reproduction and replication of their simulation study.

# Acknowledgements

## Author Contributions

KL: Project administration, Formal Analysis, Software, Validation, Visualization, Writing – original draft, Writing – review & editing. AL: Conceptualization, Project administration, Formal Analysis, Software, Writing – original draft, Writing – review & editing. UA: Formal analysis, Software, Validation, Visualizations, Writing – review & editing. JCG: Formal Analysis, Software, Validation. FJC: Software, Validation, Writing – review & editing. JLF: Formal Analysis, Software, Validation, Writing – review & editing. LH: Formal Analysis, Software, Validation. AH: Software, Validation. AKM: Software, Supervision, Validation, Writing – review & editing. LN: Writing – review & editing. RCCN: Software, Validation, Writing – review & editing. BPdV: Software, Validation, Writing – review & editing. JK: Formal Analysis, Software, Validation, Writing – review & editing. TDT: Formal Analysis, Software, Validation, Writing – review & editing. YAW: Formal Analysis, Software, Visualizations, Validation, Writing – review & editing. RHHG: Conceptualization, Software, Supervision, Validation, Writing – original draft, Writing – review & editing.

All authors approved the final submitted version of the manuscript.

## Funding

AL was funded by a personal grant from the German Academic Scholarship Foundation. RG was supported by grants from the Netherlands Organisation for Scientific Research [ZonMW, project 917.16.430] and from the Leiden University Medical Center.

TDT was funded by the National Science Foundation Graduate Research Fellowship Program: This material is based upon work supported by the National Science Foundation Graduate Research Fellowship Program under Grant No. DGE-2034835. Any opinions, findings, and conclusions or recommendations expressed in this material are those of the author(s) and do not necessarily reflect the views of the National Science Foundation. AKM, TDT, and JLF report funding provided by the National Science Foundation through Ethical and Responsible Research under award number 2024377.

**Table 1**

*Statistical simulation studies that were replicated*

| Authors (year) | Title | Number of citations in Google Scholar (March 2022) |
| --- | --- | --- |
| Austin (2011) | Optimal caliper widths for propensity score matching when estimating differences in means and differences in proportions in observational studies | 2265 |
| Brookhart, Schneeweiss and colleagues (2006) | Variable Selection for Propensity Score Models | 1911 |
| Flora & Curran (2004) | An Empirical Evaluation of Alternative Methods of Estimation for Confirmatory Factor Analysis With Ordinal Data | 2932 |
| Fritz & MacKinnon (2007) | Required Sample Size to Detect the Mediated Effect | 3784 |
| MacKinnon, Lockwood, & Williams (2004) | Confidence Limits for the Indirect Effect: Distribution of the Product and Resampling Methods | 7098 |
| Peters, Sutton and colleagues (2006) | Comparison of Two Methods to Detect Publication Bias in Meta-analysis | 1654 |
| Rhemtulla, Brosseau-Liard & Savalei (2012) | When can categorical variables be treated as continuous? | 1650 |
| Vittinghof & McCulloch (2007) | Relaxing the Rule of Ten Events per Variable in Logistic and Cox Regression | 3031 |

**Table 2**

*Experienced facilitators and barriers for replication of statistical simulation studies*

| Authors (year) | Replication facilitators | Replication barriers |
|---|---|---|
| Austin (2011) | • The magnitude of parameters for data generation was explicitly mentioned.<br>• Provided formulas made it straightforward to compute the intended performance measures. | • The computational cost of propensity-score matched samples was high. |
| Brookhart, Schneeweiss et al. (2006) | • The original article provided clear descriptions of the simulation methods.<br>• Provided formulas made it straightforward to implement the intended approach.<br>• The data-generating mechanism was depicted in a figure, which was helpful in understanding the simulation set up.<br>• The replicators were familiar with the literature in this topic. This experience could be used to assume the most likely approach for implicit decisions. | • Some of the results were presented as figures only, which hampered the exact comparison of results.<br>• Specific software implementations were not described clearly, such as how splines were fitted. Although the implementation had to be assumed, results were still replicable. |
| Flora & Curran (2004) | • The original article provided most of the theoretical information and instruction required to replicate the study.<br>• A recent article by Chalmers and Adkins (2020) provided the code for a partial replication of the same study using the SimDesign package in R statistical software. | • The exact values (tau) used to transform continuous data into ordinal data were not reported; tau values for five-category ordinal data referenced by the authors produced distributions inconsistent with the original article, but a correction reported in Chalmers and Adkins (2020) resolved the inconsistency and was used in subsequent replications. |

Fritz & MacKinnon (2007)

- The original article provided a clear and detailed description of the methods that were implemented.
- The replicators were familiar with the literature in this topic, including work by the authors of this particular study. This experience could be used to assume the most likely approach for implicit decisions.

- The original study used proprietary software EQS and Mplus, whereas the replication used open-source R statistical software.
- The links to the technical appendix of the original paper, which was described as containing example code for data generation and model estimation, were broken. The replication team was not able to locate the appendix.
- The criteria for the margin of error on power level for bootstrap simulation methods was not explicitly stated in the article. The assumed margin of error had a small influence on the replicability of results.
- One of the methods was originally written in the programming language FORTRAN and was unreadable in any open-source programming language. Instead, the RMediate package in R statistical software was used (Tofighi & MacKinnon, 2011). Although many functionalities are similar to the original method, some results could not be replicated because the version of the method written for R statistical software sometimes halts execution because of a bug (known to the authors of the method).

MacKinnon, Lockwood & Williams (2004)

- The overall structure of the simulation study was easy to glean from the original article, and the simulation conditions were explicitly laid out.

- Critical values used in one of the inferential methods included in the original simulation study were obtained from the tables in a book(Meeker et al., 1981). However, no further information was provided on how the values printed in these tables were implemented in the code for the original

- Although some were unclear, instructions were provided for implementing all methods used in the simulation study.
- The replicators were familiar with the literature in this topic, including work by the authors of this particular study. This experience could be used to assume the most likely approach for implicit decisions.

simulation study. Consequently, a similar (but different) inferential method that was available in an R package created by the original first author had to be used in the replication instead (Tofighi & MacKinnon, 2011).

- Formulas provided in the original publication to calculate certain statistics used in the inferential methods, such as t-statistics and skewness values, were unclear, and so the original formulas had to be retrieved from a book hidden behind a paywall (Manly, 1997).
- One of the methods was originally written in the programming language FORTRAN and was unreadable in any open-source programming language. Instead, the RMediate package in R statistical software was used (Tofighi & MacKinnon, 2011). Although many functionalities are similar to the original method, some results could not be replicated because the version of the method written for R statistical software sometimes halts execution because of a bug (known to the authors of the method).
- Links provided in the original publication that contained code/other information necessary for the replication attempt were broken, or they connected to a website that no longer contained the needed information. As a result, one of the inferential methods examined in the original study had to be dropped from the replication simulation.

| | | |
|---|---|---|
| Peters, Sutton et al. (2006) | • Data-generating mechanism and simulation scenarios were relatively well described in a technical report (which was not accessible in the public domain).<br>• The technical report presented complete results for all investigated simulation scenarios. | • The description of some simulation experiments was incomplete. For instance, the combination of sample size and expected event fraction could result in datasets without any events occurring, meaning that subsequential analyses could not be performed. Assumptions had to be made on how to replicate the simulation scenario.<br>• Results were presented as figures only, which hampered the exact comparison of results. |
| Rhemtulla, Brosseau-Liard & Savalei (2012) | • The original articled contained a well-structured methods section that detailed all simulation experiments in separate sections. It was clear from the descriptions how to implement the method.<br>• Descriptives for the generated data allowed for an easy check of the data generating mechanism.<br>• The original manuscript presented descriptive results of errors that had occurred. This enabled us to compare the number and type of errors that occurred in our replication to the original study.<br>• Results were presented in tables in the supplementary files. | • While error descriptives allowed for comparison of errors occurring across different software implementation, the error handling was not described in sufficient detail to evaluate the effects of the errors on the results. |
| Vittinghof & McCulloch (2007) | | • Simulation parameters were insufficiently described, hampering replication at an early stage.<br>• Data-generating procedures only contained a description of expected results without specifying the procedure to generate it. For instance, the |

correlation of a binary predictor with continuous predictors was described, but it was not indicated how the binary predictor data was generated and how the correlation with other variables was introduced.

**Supplementary File**

This supplementary file accompanies "Replicability of Simulation Studies for the Investigation of Statistical Methods: The RepliSims Project" by Kim Luijken, Anna Lohmann, Udi Alter, Juan Claramunt Gonzalez, Felix Clouth, Jessica Fossum, Lieke Hesen, Arjan Huizing, Jolien Ketelaar, Amanda Montoya, Linda Nab, Rick Nijman, Bas Penning de Vries, Tristan Tibbe, Andre Wang, and Rolf Groenwold. This file refers to the complete replication reports referred to in the results section of the main text.

All replication reports can be found on the Zenodo community "Simulation Study Replication", https://zenodo.org/communities/replisims/

Austin
- Replication report: https://doi.org/10.5281/zenodo.7474737
- Replication code: https://github.com/replisims/austin-2011

Brookhart
- Replication reports: https://doi.org/10.5281/zenodo.5075093, https://doi.org/10.5281/zenodo.6471408
- Replication codes: https://github.com/replisims/Brookhart_MA-2006, https://github.com/replisims/Brookhart-2006-FJ

Flora
- Replication report: https://doi.org/10.5281/zenodo.6460290
- Replication code: https://github.com/replisims/Flora_Curran_2004

Fritz
- Replication report: https://doi.org/10.5281/zenodo.5975519
- Replication code: https://github.com/replisims/fritz-2007

MacKinnon
- Replication report: https://doi.org/10.5281/zenodo.6460343
- Replication code: https://github.com/replisims/mackinnon-2004

Peters
- Replication report: https://doi.org/10.5281/zenodo.6595979
- Replication code: https://github.com/replisims/peters-2016

Rhemtulla
- Replication report: https://doi.org/10.5281/zenodo.6626045
- Replication code: https://github.com/replisims/rhemtulla-2012

Vittinghof
- Replication report: https://doi.org/10.5281/zenodo.5119364
- Replication code: https://github.com/replisims/vittinghoff-mcculloch-2007